\input harvmac
\input amssym

\def\ap{\alpha'}

\def\psib{\overline{\psi}}

\def\p{\partial}
\def\pb{\overline{\partial}}

\def\pb{\overline{\partial}}

\def\ct{\tilde{c}}

 \def\mb{\bar{m}}

\def\cL{{\cal L}}
\def\wb{{\overline{w}}}
\def\Ab{\overline{A}}
\def\psib{{\overline{\psi}}}
\def\Dc{{\cal D}}
\def\etah{{\hat{\eta}}}
\def\wbh{{\hat{\overline{w}}}}
\def\wh{{\hat{w}}}

\def\Ah{{\hat{A}}}
\def\Bh{{\hat{B}}}
\def\Ch{{\hat{C}}}
\def\what{\widehat} 

\baselineskip 13pt

\lref\MetsaevZX{
  R.~R.~Metsaev and A.~A.~Tseytlin,
  ``Order alpha-prime (Two Loop) Equivalence of the String Equations of Motion
  and the Sigma Model Weyl Invariance Conditions: Dependence on the Dilaton and
  the Antisymmetric Tensor,''
  Nucl.\ Phys.\  B {\bf 293}, 385 (1987).
}

\lref\BergshoeffDE{
  E.~A.~Bergshoeff and M.~de Roo,
  ``The Quartic Effective Action Of The Heterotic String And Supersymmetry,''
  Nucl.\ Phys.\  B {\bf 328}, 439 (1989).
}

\lref\AntoniadisEG{
  I.~Antoniadis, S.~Ferrara, R.~Minasian and K.~S.~Narain,
  ``R**4 couplings in M- and type II theories on Calabi-Yau spaces,''
  Nucl.\ Phys.\  B {\bf 507}, 571 (1997)
  [arXiv:hep-th/9707013].
}

\lref\CastroSD{
  A.~Castro, J.~L.~Davis, P.~Kraus and F.~Larsen,
  ``5D attractors with higher derivatives,''
  JHEP {\bf 0704}, 091 (2007)
  [arXiv:hep-th/0702072].
}

\lref\CastroHC{
  A.~Castro, J.~L.~Davis, P.~Kraus and F.~Larsen,
  ``5D Black Holes and Strings with Higher Derivatives,''
  arXiv:hep-th/0703087.
}

\lref\WittenKT{
  E.~Witten,
  ``Three-Dimensional Gravity Revisited,''
  arXiv:0706.3359 [hep-th].
}

\lref\CarlipZN{
  S.~Carlip,
  ``Conformal field theory, (2+1)-dimensional gravity, and the BTZ black
  hole,''
  Class.\ Quant.\ Grav.\  {\bf 22}, R85 (2005)
  [arXiv:gr-qc/0503022].
}

\lref\MooreYH{
  G.~W.~Moore and N.~Seiberg,
  ``Taming the Conformal Zoo,''
  Phys.\ Lett.\  B {\bf 220}, 422 (1989).
}

\lref\ElitzurNR{
  S.~Elitzur, G.~W.~Moore, A.~Schwimmer and N.~Seiberg,
  ``Remarks On The Canonical Quantization Of The Chern-Simons-Witten Theory,''
  Nucl.\ Phys.\  B {\bf 326}, 108 (1989).
}

\lref\WittenHF{
  E.~Witten,
  ``Quantum field theory and the Jones polynomial,''
  Commun.\ Math.\ Phys.\  {\bf 121}, 351 (1989).
}

\lref\HenneauxIB{
  M.~Henneaux, L.~Maoz and A.~Schwimmer,
  ``Asymptotic dynamics and asymptotic symmetries of three-dimensional
  extended AdS supergravity,''
  Annals Phys.\  {\bf 282}, 31 (2000)
  [arXiv:hep-th/9910013].
}

\lref\ReggeZD{
  T.~Regge and C.~Teitelboim,
  ``Role Of Surface Integrals In The Hamiltonian Formulation Of General
  Relativity,''
  Annals Phys.\  {\bf 88}, 286 (1974).
}

\lref\KrausVZ{
  P.~Kraus and F.~Larsen,
  ``Microscopic black hole entropy in theories with higher derivatives,''
  JHEP {\bf 0509}, 034 (2005)
  [arXiv:hep-th/0506176].
  }

\lref\KrausZM{
  P.~Kraus and F.~Larsen,
  ``Holographic gravitational anomalies,''
  JHEP {\bf 0601}, 022 (2006)
  [arXiv:hep-th/0508218].
}

\lref\HanakiPJ{
  K.~Hanaki, K.~Ohashi and Y.~Tachikawa,
  ``Supersymmetric Completion of an $R^2$ Term in Five-Dimensional
  Supergravity,''
  Prog.\ Theor.\ Phys.\  {\bf 117}, 533 (2007)
  [arXiv:hep-th/0611329].
}

\lref\SenKJ{
  A.~Sen,
  ``Stretching the horizon of a higher dimensional small black hole,''
  JHEP {\bf 0507}, 073 (2005)
  [arXiv:hep-th/0505122].
}

\lref\HMM{ J.~A.~Harvey, R.~Minasian and G.~W.~Moore, ``Non-abelian
tensor-multiplet anomalies,''
 JHEP {\bf 9809}, 004 (1998)
  [arXiv:hep-th/9808060].
}

\lref\StromStrings{
A.~Strominger, talk at STRINGS '07 (Madrid), june 25, 2007.
}

\lref\Strompaper{
J.~Lapan, A.~Simons and A.~Strominger, to appear. As advertized in
\StromStrings.
}

\lref\KnizhnikWC{
  V.~G.~Knizhnik,
  ``Superconformal algebras in two dimensions,''
  Theor.\ Math.\ Phys.\  {\bf 66}, 68 (1986)
  [Teor.\ Mat.\ Fiz.\  {\bf 66}, 102 (1986)].
}
\lref\BershadskyMS{
  M.~A.~Bershadsky,
  ``Superconformal algebras in two dimensions with arbitrary N,''
  Phys.\ Lett.\  B {\bf 174}, 285 (1986).
}
\lref\DefeverTF{
  F.~Defever, W.~Troost and Z.~Hasiewicz,
  ``Superconformal algebras with quadratic nonlinearity,''
  Phys.\ Lett.\  B {\bf 273}, 51 (1991).
}
\lref\FradkinKM{
  E.~S.~Fradkin and V.~Y.~Linetsky,
   ``Results Of The Classification Of Superconformal Algebras In
  Two-Dimensions,''
  Phys.\ Lett.\  B {\bf 282}, 352 (1992)
   [arXiv:hep-th/9203045].
  ``Classification Of Superconformal Algebras With Quadratic Nonlinearity,''
  arXiv:hep-th/9207035.
    ``Classification Of Superconformal And Quasisuperconformal Algebras In
  Two-Dimensions,''
  Phys.\ Lett.\  B {\bf 291}, 71 (1992);
}

\lref\FradkinGJ{
  E.~S.~Fradkin and V.~Y.~Linetsky,
  ``An Exceptional N=8 superconformal algebra in two-dimensions associated with
  F(4),''
  Phys.\ Lett.\  B {\bf 275}, 345 (1992).
}

\lref\GoddardWV{
  P.~Goddard and A.~Schwimmer,
  ``FACTORING OUT FREE FERMIONS AND SUPERCONFORMAL ALGEBRAS,''
  Phys.\ Lett.\  B {\bf 214}, 209 (1988).
}

\lref\SchoutensTG{
  K.~Schoutens,
  ``Representation Theory For A Class Of SO(N) Extended Superconformal Operator
  Algebras,''
  Nucl.\ Phys.\  B {\bf 314}, 519 (1989).
}

\lref\NahmTG{
  W.~Nahm,
  ``Supersymmetries and their representations,''
  Nucl.\ Phys.\  B {\bf 135}, 149 (1978).
}

\lref\BalasubramanianRE{
  V.~Balasubramanian and P.~Kraus,
  ``A stress tensor for anti-de Sitter gravity,''
  Commun.\ Math.\ Phys.\  {\bf 208}, 413 (1999)
  [arXiv:hep-th/9902121].
}

\lref\EmparanPM{
  R.~Emparan, C.~V.~Johnson and R.~C.~Myers,
  ``Surface terms as counterterms in the AdS/CFT correspondence,''
  Phys.\ Rev.\  D {\bf 60}, 104001 (1999)
  [arXiv:hep-th/9903238].
  }

\lref\KrausDI{
  P.~Kraus, F.~Larsen and R.~Siebelink,
  ``The gravitational action in asymptotically AdS and flat spacetimes,''
  Nucl.\ Phys.\  B {\bf 563}, 259 (1999)
  [arXiv:hep-th/9906127].
}

\lref\PapadimitriouII{
  I.~Papadimitriou and K.~Skenderis,
  ``Thermodynamics of asymptotically locally AdS spacetimes,''
  JHEP {\bf 0508}, 004 (2005)
  [arXiv:hep-th/0505190].
}

\lref\HansenWU{
  J.~Hansen and P.~Kraus,
  ``Generating charge from diffeomorphisms,''
  JHEP {\bf 0612}, 009 (2006)
  [arXiv:hep-th/0606230].
}

\lref\GiveonNS{
  A.~Giveon, D.~Kutasov and N.~Seiberg,
  ``Comments on string theory on AdS(3),''
  Adv.\ Theor.\ Math.\ Phys.\  {\bf 2}, 733 (1998)
  [arXiv:hep-th/9806194].
}
\lref\KutasovXU{
  D.~Kutasov and N.~Seiberg,
  ``More comments on string theory on AdS(3),''
  JHEP {\bf 9904}, 008 (1999)
  [arXiv:hep-th/9903219].
}
\lref\GiveonPR{
  A.~Giveon and D.~Kutasov,
  ``Fundamental strings and black holes,''
  JHEP {\bf 0701}, 071 (2007)
  [arXiv:hep-th/0611062].
}
\lref\KutasovZH{
  D.~Kutasov, F.~Larsen and R.~G.~Leigh,
  ``String theory in magnetic monopole backgrounds,''
  Nucl.\ Phys.\  B {\bf 550}, 183 (1999)
  [arXiv:hep-th/9812027].
}
\lref\GiddingsWN{
  S.~B.~Giddings, J.~Polchinski and A.~Strominger,
  ``Four-dimensional black holes in string theory,''
  Phys.\ Rev.\  D {\bf 48}, 5784 (1993)
  [arXiv:hep-th/9305083].
}
\lref\JohnsonJW{
  C.~V.~Johnson,
  ``Exact models of extremal dyonic 4-D black hole solutions of heterotic
  string theory,''
  Phys.\ Rev.\  D {\bf 50}, 4032 (1994)
  [arXiv:hep-th/9403192].
}

\lref\FradkinBZ{
  E.~S.~Fradkin and V.~Y.~Linetsky,
  ``Results Of The Classification Of Superconformal Algebras In
  Two-Dimensions,''
  Phys.\ Lett.\  B {\bf 282}, 352 (1992)
  [arXiv:hep-th/9203045].
}

\lref\BinaBM{
  B.~Bina and M.~Gunaydin,
  ``Real forms of non-linear superconformal and quasi-superconformal  algebras
  and their unified realization,''
  Nucl.\ Phys.\  B {\bf 502}, 713 (1997)
  [arXiv:hep-th/9703188].
}

\lref\BanadosEY{
  M.~Banados, O.~Chandia and A.~Ritz,
  ``Holography and the Polyakov action,''
  Phys.\ Rev.\  D {\bf 65}, 126008 (2002)
  [arXiv:hep-th/0203021].
}

\lref\VanProeyenNI{
  A.~Van Proeyen,
  ``Tools for supersymmetry,''
  arXiv:hep-th/9910030.
}

\lref\deBoerIP{
  J.~de Boer,
  ``Six-dimensional supergravity on S**3 x AdS(3) and 2d conformal field
  theory,''
  Nucl.\ Phys.\  B {\bf 548}, 139 (1999)
  [arXiv:hep-th/9806104].
}

\lref\IshimotoHD{
  Y.~Ishimoto,
  ``Classical Hamiltonian reduction on D(2|1,alpha) Chern-Simons gauge  theory
  and large N = 4 superconformal symmetry,''
  Phys.\ Lett.\  B {\bf 458}, 491 (1999)
  [arXiv:hep-th/9808094].
}

\lref\ItoVD{
  K.~Ito,
  ``Extended superconformal algebras on AdS(3),''
  Phys.\ Lett.\  B {\bf 449}, 48 (1999)
  [arXiv:hep-th/9811002].
}

\lref\BowcockBM{
  P.~Bowcock,
  ``Exceptional Superconformal Algebras,''
  Nucl.\ Phys.\  B {\bf 381}, 415 (1992)
  [arXiv:hep-th/9202061].
}

\lref\GukovFH{
 S.~Gukov, E.~Martinec, G.~W.~Moore and A.~Strominger,
 ``An index for 2D field theories with large N = 4 superconformal  symmetry,''
arXiv:hep-th/0404023.}

\lref\LapanJX{
  J.~M.~Lapan, A.~Simons and A.~Strominger,
  ``Nearing the Horizon of a Heterotic String,''
  arXiv:0708.0016 [hep-th].
}

\lref\BrownNW{
  J.~D.~Brown and M.~Henneaux,
  ``Central Charges in the Canonical Realization of Asymptotic Symmetries: An
  Example from Three-Dimensional Gravity,''
  Commun.\ Math.\ Phys.\  {\bf 104}, 207 (1986).
}

\lref\polch{Polchinski}

\lref\yellow{
  P.~Di Francesco, P.~Mathieu and D.~Senechal,
  ``Conformal Field Theory,''
{\it  New York, USA: Springer (1997) 890 p}
}

\lref\DabholkarGP{
  A.~Dabholkar and S.~Murthy,
  ``Fundamental Superstrings as Holograms,''
  arXiv:0707.3818 [hep-th].
}

\lref\JohnsonDU{
  C.~V.~Johnson,
  ``Heterotic Coset Models of Microscopic Strings and Black Holes,''
  arXiv:0707.4303 [hep-th].
}

\lref\KrausNB{
  P.~Kraus and F.~Larsen,
  ``Partition functions and elliptic genera from supergravity,''
  JHEP {\bf 0701}, 002 (2007)
  [arXiv:hep-th/0607138].
}

\lref\LopesCardosoWT{
  G.~Lopes Cardoso, B.~de Wit and T.~Mohaupt,
  ``Corrections to macroscopic supersymmetric black-hole entropy,''
  Phys.\ Lett.\  B {\bf 451}, 309 (1999)
  [arXiv:hep-th/9812082].
}

\lref\DabholkarYR{
  A.~Dabholkar,
  ``Exact counting of black hole microstates,''
  Phys.\ Rev.\ Lett.\  {\bf 94}, 241301 (2005)
  [arXiv:hep-th/0409148].
}

\lref\SenDP{
  A.~Sen,
  ``How does a fundamental string stretch its horizon?,''
  JHEP {\bf 0505}, 059 (2005)
  [arXiv:hep-th/0411255].
}

\lref\HubenyJI{
  V.~Hubeny, A.~Maloney and M.~Rangamani,
  ``String-corrected black holes,''
  JHEP {\bf 0505}, 035 (2005)
  [arXiv:hep-th/0411272].
}

\Title{\vbox{\baselineskip12pt
}}
{\vbox{\centerline
{Fundamental Strings, Holography,}\medskip\vbox{\centerline {and Nonlinear Superconformal Algebras}}} } 
\centerline{
Per Kraus$^{\spadesuit}$\foot{pkraus@ucla.edu},  Finn
Larsen$^\dagger$\foot{larsenf@umich.edu}, and Akhil Shah$^{\spadesuit}$\foot{akhil137@ucla.edu}
}
\bigskip

\centerline{${}^{\spadesuit}$\it{Department of Physics and
Astronomy, UCLA,}}\centerline{\it{ Los Angeles, CA 90095-1547,
USA.}}\vskip.2cm \centerline{${}^\dagger$\it{Department of Physics
and Michigan Center for Theoretical Physics,
}} \centerline{\it{University of Michigan, Ann
Arbor, MI 48109-1120, USA.}}

\baselineskip15pt

\vskip .3in

\centerline{\bf Abstract}
We discuss aspects of holography in the $AdS_3\times S^p$ near string geometry
of a collection of straight fundamental heterotic strings. We use anomalies
and symmetries to determine general features of the dual CFT.  The symmetries suggest
the appearance of nonlinear superconformal algebras, and we show how these arise
in the framework of holographic renormalization methods.  The nonlinear algebras
imply intricate formulas for the central charge, and we show that in the bulk these correspond to
an infinite series of quantum gravity corrections.      We also makes some comments on the
worldsheet $\sigma$-model for strings on
$AdS_3\times S^2$, which is the holographic dual geometry of  parallel heterotic
strings in five dimensions.

\Date{August, 2007}
\baselineskip14pt

\newsec{Introduction}
One can attempt to describe the  fundamental string in supergravity by looking for a solution
with the same symmetries and charges as a fundamental string.  For example, in the
classical two-derivative approximation,  $N$ fundamental heterotic strings in
5D are described by the geometry
\eqn\aaa{
ds^2_{5E} = H^{-1/3} (-dt^2 + dy^2) + H^{2/3}(dr^2 + r^2 d\Omega^2)~,
}
where
\eqn\aae{
H = 1 + {\sqrt{\alpha^\prime} N\over 2r}~.
}
Away from the origin, this solution captures the nontrivial backreacted geometry
produced by the string,  but at the origin the curvature diverges, signaling the need to
retain higher derivative string theory corrections.     The leading order corrections
in heterotic string theory arise at four-derivatives ($R^2$), and are known in detail \refs{\MetsaevZX,\BergshoeffDE,\AntoniadisEG,\HanakiPJ}.
Keeping just corrections of this order,  the fundamental string
 solution (given in \refs{\CastroSD,\CastroHC})
then takes the form \aaa,  with $H$ interpolating from the
large $r$ behavior  \aae\ to the  $r\rightarrow 0$ near string  limit
\eqn\aab{
H\to {\ell^3\over r^3}~,\quad \ell = \sqrt{\alpha^\prime\over 2}~.
}
The corresponding near string geometry is $AdS_3\times S^2$, a smooth space. The
string coupling at the origin vanishes in the leading approximation,
but the $R^2$-corrections stabilize it at
\eqn\aac{
g^{\rm het}_5 = 2^{1/4} N^{-1/2}~.
}
The string coupling can be made arbitrarily small for large $N$, so it is meaningful to use the near string solution as the target space for a string theory $\sigma$-model.   Such a description is
also facilitated by the absence of Ramond sector fields.
The resulting theory is
interesting because, according to AdS/CFT, it provides a dual description of the
original source strings. In other words, the bulk theory is the holographic dual of $N$
fundamental heterotic strings.

There are a number of obstacles to surmount in order to make this story
precise.  In terms of applying  higher derivative corrections to the supergravity action,
a general problem with any solution of order string string scale is that there
is no small expansion parameter, and so one generally expects an infinite
series of corrections that are all of the same order.   In a similar vein, there are
field redefinition ambiguities to contend with, since, for example,  $g_{\mu\nu}$
and $\alpha' R_{\mu\nu}$ are of the same order. What makes the present situation
special is that symmetries and anomalies in the near horizon region are so powerful
that some features of the solution are protected from corrections beyond the four-derivative order.
An important example is that the spacetime central charges can be determined exactly
in this manner, which in turn leads to precise results for black hole entropy.

The next issue appears when we consider the symmetries of the
near horizon geometry of fundamental strings.   If we extend our discussion to the case
of a straight fundamental string in $p+3$ dimensions, then we expect the existence
of an AdS$_3 \times S^p$ near horizon geometry.     The presence of the AdS$_3$
factor implies the existence of left and right moving Virasoro algebras \BrownNW.
Further, we expect the near horizon
geometry to preserve 16 supercharges, due to the usual near horizon enhancement.
Based on the worldsheet structure of the heterotic string, we thus expect a $(0,8)$
theory with $SO(p+1)$ acting as an R-symmetry.  The puzzle is that in the standard
list of superconformal algebras one has at most 4 right moving supercurrents.

There is at least one potential resolution of this puzzle: there exist {\it nonlinear  superconformal
algebras} with the required symmetries \refs{\KnizhnikWC,\BershadskyMS}.
These algebras contain just the stress tensor, R-symmetry currents, and
supercurrents,  and are classified according to their finite dimensional supergroups.
The algebras are nonlinear in that bilinears of the R-currents appear in
the OPEs of the supercurrents. These nonlinear algebras have been studied from the algebraic
perspective \refs{\SchoutensTG,\DefeverTF,\FradkinKM,\FradkinGJ,\BowcockBM,\BinaBM} but have not found many
physical applications.   One place where they are known to appear is as the asymptotic
symmetry algebra of AdS$_3$ supergravity based on arbitrary supergroups, as was
noted in \refs{\deBoerIP,\ItoVD}, and shown in detail in \HenneauxIB.
Since they naturally appear  in the context of
AdS$_3$ supergravity, we are led to conjecture that the symmetry algebras of our
heterotic string solutions are precisely these nonlinear superconformal algebras.

This proposal has a number of interesting implications, of which we now mention one  (there are
also some  puzzles, as we discuss later on).
The algebras are each parameterized by a parameter $k$, identified with the level of
the R-symmetry current algebra.   Jacobi identities give a formula for the central charge in
terms of $k$ that has the structure $c(k) \sim k + k^0 + {1 \over k} + \ldots$.   By contrast,
the ordinary superconformal algebras have $c \sim k$.      As we'll show explicitly, $k$ is
proportional to the number of heterotic strings $N$.   From \aac\ we  see that the
expansion in $1/k$ is equivalent to an expansion in  $g_s^2$, i.e. in Newton's constant.
In other words, we find that the Jacobi identities of the nonlinear algebras imply an infinite
series of quantum gravity corrections to the central charges.   Using standard methods,
we can translate this result into quantum gravity corrections to the entropy of  black
holes in this theory.  We find it remarkable that (under our assumptions) these corrections
are determined algebraically.

This paper is aimed at collecting some results  and observations related to the
holographic duals of heterotic strings, mainly based on the hypothesis that they
are governed by the nonlinear superconformal algebras.
We start by discussing the explicit  solution for the geometry near
$N$ fundamental strings in 5D  when $R^2$-corrections are taken into account.
For this we must dualize previous results \refs{\CastroSD,\CastroHC}
to the heterotic frame. We next revisit the symmetry and anomaly inflow arguments
from \refs{\HMM,\KrausVZ, \KrausZM} and generalize them to $AdS_3\times S^p$
solutions for all $p$.   After a review of the salient aspects of the nonlinear algebras,
we give results for the exact central charges including quantum corrections.
For example, we find results for the central charges for geometries like
$AdS_3\times S^7$ that have not yet been found
as solutions to supergravity with higher derivative terms.

In the next several sections of the paper we study the symmetry aspects of the holographic duality.
We set up the holographic renormalization formalism needed to regulate and
interpret infrared divergences in AdS \refs{\BalasubramanianRE,\EmparanPM,\KrausDI,\PapadimitriouII}.   We show how the nonlinear algebras follow from
systematic application of AdS/CFT, reproducing the results of \HenneauxIB.   This
classical treatment only  gives the large $k$ limit of the algebra (the Poisson bracket algebra),
but we explain how the full quantum algebra arises from the bulk point of view.  The quantum
corrrections to the central charge can be understood in the more familar context of the
formula for the Sugawara central charge, and we review the corresponding AdS$_3$
side of the story.

In the last section of the paper we make some comments on  the string theory $\sigma$-model on
$AdS_3\times S^2$. We review some of the standard results on string theory in
$AdS_3$ and find that  the simplest candidates reproduce most, but not all, of the
features expected from the spacetime point of view. It is an interesting problem
to find a workable worldsheet theory with the correct properties.

\medskip
\noindent
{\bf Note:}
As  this work was being completed, a talk by Strominger at Strings 2007 
reported results which overlap with some of those presented here; in particular,
the relevance of the nonlinear superconformal algebras.  More recently, 
the papers  \refs{\DabholkarGP,\JohnsonDU,\LapanJX} have appeared,
all dealing with aspects of holography for fundamental strings.

\newsec{Nonsingular heterotic string solutions in 5D}

In \refs{\CastroSD,\CastroHC} nonsingular supergravity solutions
were found representing a collection of straight heterotic strings in 5D.
The supersymmetric completion of certain $R^2$-terms \HanakiPJ\ were included,
as they must be in order to avoid naked singularities. The explicit solutions asymptote
between 5D Minkowski spacetime and a near horizon AdS$_3\times S^2$ region.
If we  replace AdS$_3$ by a BTZ black hole and dimensionally reduce
along the horizon direction we recover the small 4D black holes studied in 
\refs{\LopesCardosoWT,\DabholkarYR,\SenDP,\HubenyJI}.

The solution in \refs{\CastroSD,\CastroHC} is presented in terms of
M-theory compactified on $K3\times T^2$, with $N$ M5-branes
wrapped on $K3$. In this description the horizon attractor value
for the moduli was found to give the volume
Vol$(T^2)=M^1(2\pi l_P)^2$ with
the modulus $M^1= 2^{-1/3}N^{2/3}$ and $l_P$ is the 11D Planck
length. In type IIA variables\foot{The string units are
given as usual as $l_P = g^{1/3}_s l_s$, $R_{11}=g_s l_s = g_s^{2/3}l_P$.
This gives $g_s = (R_{11}/l_P)^{3/2} = (M^1)^{3/4}$. } the corresponding string coupling
is
\eqn\aag{
g^{IIA}_s = 2^{-{1/4}}N^{1/2}~.
}
We choose to fix the total volume of $K3\times T^2$ to be the unit volume
$(2\pi l_P)^6$, so
\eqn\aah{
{\rm Vol}(K3)= (M^1)^{-1}(2\pi l_P)^4= (2\pi l_s)^4~,
}
where $l_s = \sqrt{\ap}$.

In this paper we want to analyze physics in the heterotic string frame. For this we use
six-dimensional
IIA-heterotic duality, under which the dilaton transforms as
$e^{\phi_6^{\rm het}} = e^{-\phi_6^{\rm IIA}}$,  and the Einstein frame metric is invariant.
To proceed, we use the standard relation between Einstein and string frame metrics,
\eqn\ab{g^E_{\mu\nu}=e^{-{4\phi_D \over D-2}}g^S_{\mu\nu}~,}
as well as the redefinition of the  dilaton under dimensional reduction,
\eqn\ac{ e^{-2\phi_{D-p}}  =  \sqrt{  {\rm det} g^S_{mn}}  e^{-2\phi_D} ~, }
where $g^S_{mn}$ is the string frame metric of the compact space.

On the IIA side we find $e^{\phi_6^{IIA}} = e^{\phi_{10}^{IIA} } = g_s^{IIA}$, and therefore
the heterotic string coupling is
\eqn\aai{
g^{\rm het}_s = e^{\phi_6^{\rm het}}= 2^{1/4} N^{-{1/2}}~.
}
The heterotic string is therefore weakly coupled for large $N$.

Next, we consider the length scales of the geometry as measured in the heterotic string frame.
Using the invariance of the 6D Einstein metric and \ab, the conversion factor is
\eqn\azs{ L^{\rm het} = {1 \over g_s^{IIA} } L^{IIA}~,}
where each length is measured with respect
to the corresponding string length.

The heterotic dual is compactified to 6D on $T^4$ and there is
an additional $S^1$ bringing the theory to 5D.      On the IIA side this circle has size
$ \sqrt{M^1} (2\pi l_P) = \sqrt{M^1}(g_s^{IIA})^{1\over 3} (2\pi l_s )$.    Applying the
conversion factor \azs, as well as $M^1 =  (g_s^{IIA})^{4\over 3}$, we find the size on the
heterotic side to be
\eqn\azt{ 2\pi R_5^{\rm het} = 2\pi l_s~.}
This is precisely the  self-dual radius, which  is interesting because it makes
enhanced symmetry possible.

Let us now turn to the 5D part of the geometry. In the IIA variables the
AdS$_3$ and $S^2$ radii were determined to have sizes \refs{\CastroSD,\CastroHC}
\eqn\aak{ \ell^{IIA}_S = {1\over 2}\ell^{IIA}_A =  \left( {1\over 4} N\right)^{1/3}l_P= {1 \over \sqrt{2}} g_s^{IIA} l_s~.}
The heterotic geometry is also AdS$_3\times S^2$ but the overall scale is
different.  From \azs\ we find

\eqn\jj{
\ell_S^{\rm het} = {1\over 2}\ell^{\rm het}_A = {1\over\sqrt{2}}l_s~.
}
This is of string scale  independent of the number of heterotic strings $N$.
The string scale size of the geometry is in accord with the
general scaling arguments of Sen (e.g. \SenKJ).

One might wonder why we bothered to keep track of the precise numerical
factors in the above, given that, as discussed in the introduction, the geometry is in principle
subject to corrections from higher derivative terms as well as field redefinitions.
The motivation for this is that our results might turn out to be unexpectedly robust.
The results for the geometry were derived in the context of the supersymmetric action of \HanakiPJ\
in which the supersymmetry variations are uncorrected by higher derivative terms.
This principle removes the field redefinition ambiguity (or rather defines a preferred choice
of fields).   Also, we expect that the solution actually preserves $16$ supersymmetries,
not all of which are manifest, and it could be that some of our results are fixed by this large
amount of supersymmetry.

\newsec{Chern-Simons terms for heterotic string solutions}

In this section we explain the significance of Chern-Simons terms
for AdS$_3 \times S^p \times T^{7-p}$ solutions to heterotic string
theory. We assume a constant dilaton and a uniform H-flux on AdS$_3$
carrying the charge of $N$ heterotic strings.  The only explicitly known solutions of this type are the $p=2$
solutions discussed in the last section. We can nevertheless try to
anticipate some features of solutions for general $p$.

Since these geometries include an AdS$_3$ factor, there is a
corresponding asymptotic symmetry algebra containing left and
right moving Virasoro algebras with central charges $c_{L,R}$ \BrownNW.
Further, isometries of the $S^p$ yield an $SO(p+1)$ current
algebra at some level $k$.  We would like to relate these
parameters to the number of heterotic strings $N$.

To do this we first recall that anomaly inflow \HMM\ relates the central charges
and  current algebra levels to the coefficients of bulk Chern-Simons terms
on AdS$_3$ (we review this in section 5.)  The precise relation derived in \KrausZM\ lets us write
the Chern-Simons couplings on AdS$_3$ as
\eqn\ba{S_{\rm CS} = {c_L - c_R \over 96\pi} \int\!\Omega(\omega)
+ {k\over 8\pi x_v} \int\!\Omega(A)~,}
where $\omega$ is the spin-connection,
$A$ is the $SO(p+1)$ connection, and $\Omega$ is the
Chern-Simons 3-form
\eqn\baa{\Omega(\omega) ={\rm Tr}(\omega d\omega +{2 \over
3}\omega^3)~,\quad \Omega(A)={\rm Tr}_v(A dA +{2 \over 3}A^3)~.}
In \ba\ $x_v$ denotes the Dynkin index of the vector representation of
$SO(p+1)$; it is equal to 2 for $SO(3)$ and $1$ for the other cases.

Our strategy will be to extract terms of the form \ba\ from the complete
spacetime action reduced to AdS$_3$ in the presence of $N$ units
of flux. We emphasize again that the geometries in question are
of order string scale for all $N$, and so we cannot neglect $\ap$-corrections
when doing this. We write the bosonic sector of the heterotic string action including all
higher derivative terms schematically as
\eqn\jm{\eqalign{ S&= {1 \over 2 \kappa^2_{10} }\int\! d^{10}x
\sqrt{-g}~  \cL(g_{MN}, \phi, H, A_{\rm het})~.  }}
Here $\cL$ is a function of the metric, dilaton, 3-form field strength,
Yang-Mills fields, and their derivatives.  In heterotic string theory
the 3-form field strength is determined by anomaly
cancellation to be of the form
\eqn\jma{ H = dB+ {\ap \over 4} \Omega(\omega)- {\ap \over 4}
\Omega(A_{\rm het})~.}
 The Yang-Mills fields
$A_{\rm het}$   will be set to zero for the time being; we comment on their
role at the end of the calculation.

We need an expression for the number $N$ of heterotic strings.
Gauss' law for the flux $\Pi^{MNP} = {\p \cL \over \p H_{MNP}}$ states
that the surface integral $\int\! ^\star \Pi$ is independent of the radial
location of the surface. In the near horizon region our {\it ansatz} gives
\eqn\jmd{\Pi^{MNP}=-{1 \over 6} Q \epsilon^{MNP}~,}
where $\epsilon^{MNP}$ is the volume form of AdS$_3$ and $Q$ is a parameter
proportional to the number of fundamental strings.
The solutions we consider asymptote to a flat region where
all higher derivative terms are negligible. There the conserved charge
can be derived from  the usual two-derivative action
\eqn\jmb{S_2 = {1 \over 2\kappa_{10}^2}\int\! d^{10}x \sqrt{-g}
e^{-2\phi}\left[ R + 4 \p^M \phi \p_M \phi -{1 \over 12}
H^{MNP}H_{MNP}\right]~,}
which gives the standard expression for $N$
\eqn\jmc{ N = {2\pi \ap \over 2\kappa_{10}^2} \int_\infty\!~{^\star} \!H~,
 }
with the integral evaluated over the  asymptotic $S^p \times
T^{7-p}$. Flux conservation and \jmd\ then relates $N$ to the near horizon data as
\eqn\bbb{N = {2\pi \ap V_7\over 2\kappa_{10}^2} \int \!~{^\star} \Pi=
{2\pi \ap Q V_7 \over 2\kappa_{10}^2  }~,}
where $V_7$ is the volume of $S^p \times T^{7-p}$.

The 10 dimensional origin of the Chern-Simons terms on AdS$_3$ is
seen from \jma.  In particular, we are just interested in the
terms in the action linear in $\Omega_{MNP}$.     The relevant term is
\eqn\jbm{\eqalign{ S_{CS} &={1 \over 2\kappa_{10}^2 }\int\!
d^{10}x \sqrt{-g}~  {\ap \over 4} {\p \cL \over \p H_{MNP} }
\Omega_{MNP}~.
}}
We can now use \jmb, \bbb, and the definition of $\Pi$ to express this as
\eqn\jsb{S_{CS} = {N \over 8\pi} \int_{AdS_3}\! \Omega~. }

We have now almost achieved our goal of extracting the Chern-Simons
terms \ba\ from the general action \jm. The only missing ingredient is
that so far we just considered pure AdS$_3 \times S^p \times T^{7-p}$.
In this case the $SO(p+1)$ connection vanishes, and so in \jsb\ we
have $\Omega =\Omega(\omega)$. To turn on a $SO(p+1)$ connection
along with the gravitational terms we replace the sphere metric
$ds^2 = \sum_{i=1}^{p+1} dy^i dy^i$  (with
$\sum_{i=1}^{p+1}y^i y^i=1$)  by $ds^2 = \sum_{i=1}^{p+1}
(dy^i-A^{ij}y^j) (dy^i-A^{ik}y^k)$ (for details on this see \HansenWU).
The Chern-Simons terms
then become
\eqn\jsc{S_{CS} = {N \over 8\pi} \int_{AdS_3}\! \Omega(\omega)+{N
\over 8\pi} \int_{AdS_3}\! \Omega(A) ~. }
Comparing with \ba\ we finally read off the relation between the central
extensions and the number of fundamental strings
\eqn\jsd{ c_L-c_R = 12N~,\quad k = \left\{ \matrix{2N~,\quad &p=2\cr N~,\quad & p>2}\right. }
This is the result we wanted to establish. The result is particularly useful when  the level $k$ is
related by supersymmetry to one of the central charges, since then the
equations determine the central charges $c_{L,R}$ separately. For example, large 5D
black strings preserve a right-moving ${\cal N}=4$ algebra with $SO(3)$ R-symmetry
and so $c_R = 6k= 12N$ \KrausVZ. However, the fundamental heterotic string solutions are expected
to have more supersymmetry, and hence we can not assume this relation in general.
This is the topic of the next section.

Our derivation gives the exact answer for $c_L-c_R$ and $k$
provided that the complete contribution to the Chern-Simons terms
comes from \jma\ in this manner.  To establish this completely
we would have to carry out a complete Kaluza-Klein
reduction of the ten dimensional action down to three dimensions,
which is obviously not possible given our general starting point.
While it is possible that additional Chern-Simons terms are
generated by the details of the Kaluza-Klein reduction, we view
this as unlikely, and will henceforth assume that \jsd\ are the
correct expressions.

Returning to the $\Omega(A_{\rm het})$ term in \jma, we can repeat the previous
derivation and deduce the Chern-Simons term for the heterotic gauge fields
\eqn\jse{ S_{CS}= -{N\over 8\pi} \int_{AdS_3} \! \Omega(A_{\rm het}).}
For example, for the $SO(32)$ heterotic string, we then find a left moving $SO(32)$
current algebra at level $k=N$.\foot{Note that the relative minus sign compared to \jsc\ is what tell us
that this current algebra is left moving.}      For  compactification on $T^{7-p}$ we also have
$U(1)^{7-p} \times U(1)^{7-p}$ gauge fields.  At a generic point in moduli space, these
combine with the unbroken $SO(32)$ gauge fields in a  $SO(23-p,7-p)$ invariant
fashion.   That is to say, we find a   signature $(23-p, 7-p)$  spectrum of $U(1)$ currents.

\newsec{Nonlinear superconformal algebras and exact central
charges}

In this section we  try  to identify the relevant superconformal
algebras governing our spacetimes.  One benefit of being able to determine the correct
algebra is that  this will determine the quantum corrections to
the central charges in \jsd.

As before, we assume the existence of AdS$_3 \times S^p \times T^{7-p}$
solutions and use general principles to anticipate their properties.
Specifically, we expect the solution to preserve $16$ supersymmetries
due to the usual near horizon supersymmetry enhancement, and these
should all be rightmoving supersymmetries in the boundary superconformal algebra.
The global symmetry group should therefore contain one
of the four supergroups with $16$ supercharges:
\bigskip
\vbox{
$$\vbox{\offinterlineskip
\hrule height 1.1pt
\halign{&\vrule width 1.1pt#
&\strut\quad#\hfil\quad&
\vrule width 1.1pt#
&\strut\quad#\hfil\quad&
\vrule width 1.1pt#\cr
height3pt
&\omit&
&\omit&
\cr
&\hfil Supergroup&
&\hfil R-symmetry &
\cr
height3pt
&\omit&
&\omit&
\cr
\noalign{\hrule height 1.1pt}
height3pt
&\omit&
&\omit&
\cr
&\hfil $OSp(8|2;R)$ &
&\hfil $SO(8)$&
\cr
height3pt
&\omit&
&\omit&
\cr
\noalign{\hrule}
height3pt
&\omit&
&\omit&
\cr
&\hfil $F(4)$&
&\hfil $Spin(7)$&
\cr
height3pt
&\omit&
&\omit&
\cr
\noalign{\hrule}
height3pt
&\omit&
&\omit&
\cr
&\hfil $SU(1,1|4)$&
&\hfil $U(1)\times SU(4)$&\cr
height3pt
&\omit&
&\omit&
\cr
\noalign{\hrule}
height3pt
&\omit&
&\omit&
\cr
&\hfil $OSp(4^*|4) $&
&\hfil $SU(2)\times Sp(4)$&
\cr
height3pt
&\omit&
&\omit&
\cr
}\hrule height 1.1pt
}
$$
}
\centerline{\sl Table 1: Supergroups with $16$ supersymmetries, and their R-symmetry groups.}

\bigskip
We also expect the R-symmetry group to include an $SO(p+1)$ factor from isometries of
the sphere and this further helps identifying the candidate group in specific examples.

We want to determine the local symmetries of the dual boundary theory for a given
supergroup. We can immediately rule out the ordinary superconformal algebras,
since these have at most $8$ supersymmetries \NahmTG. It is then natural to
turn to the ``nonlinear superconformal algebras" \refs{\KnizhnikWC,\BershadskyMS},
since these can have additional supersymmetries with large R-symmetries, and furthermore
arise as the asymptotic symmetry algebras of AdS$_3$ supergravity
based on the corresponding supergroups \HenneauxIB. The nonlinearity refers to
the fact that the OPE of two supercurrents includes bilinears of the R-symmetry
currents. The nonlinear superconformal algebras have been classified as \refs{\KnizhnikWC,\BershadskyMS,\DefeverTF,\FradkinKM,\FradkinGJ,\BowcockBM}:
\bigskip
\vbox{
$$\vbox{\offinterlineskip
\hrule height 1.1pt
\halign{&\vrule width 1.1pt#
&\strut\quad#\hfil\quad&
\vrule width 1.1pt#
&\strut\quad#\hfil\quad&
\vrule width 1.1pt#
&\strut\quad#\hfil\quad&
\vrule width 1.1pt#\cr
height3pt
&\omit&
&\omit&
&\omit&
\cr
&\hfil Superalgebra&
&\hfil Supercurrent rep. $\rho$ &
&\hfil central charge &
\cr
height3pt
&\omit&
&\omit&
&\omit&
\cr
\noalign{\hrule height 1.1pt}
height3pt
&\omit&
&\omit&
&\omit&
\cr
&\hfil $\what{OSp}(n|2;R)$ &
&\hfil ${\bf n}$ of $SO(n)$&
&\hfil ${k(6k+n^2-10)\over 2(k+n-3)}
$&
\cr
height3pt
&\omit&
&\omit&
&\omit&
\cr
\noalign{\hrule}
height3pt
&\omit&
&\omit&
&\omit&
\cr
&\hfil $\what{SU}(1,1|n)_{n\neq 2}$&
&\hfil ${\bf n}\oplus {\bf {\bar n}}$ of $U(n)$&
&\hfil $  {3k(n+2k)+(n-1)(1+(n+1)k)\over k+n-1}$&
\cr
height3pt
&\omit&
&\omit&
&\omit&
\cr
\noalign{\hrule}
height3pt
&\omit&
&\omit&
&\omit&
\cr
&\hfil $\what{SU}(1,1|2)/U(1)$&
&\hfil ${\bf 2}\oplus {\bf {\bar 2}}$ of $SU(2)$&
&\hfil $6k$&
\cr
height3pt
&\omit&
&\omit&
&\omit&
\cr
\noalign{\hrule}
height3pt
&\omit&
&\omit&
&\omit&
\cr
&\hfil $\what{OSp}(4^*|2m)$&
&\hfill $({\bf 2},{\bf 2m})$ of $SU(2)\times Sp(2m)$ &
&\hfil $-{[6k-(2m+1)(m-2)][k+m+2]-6k \over k-m+2}$&
\cr
height3pt
&\omit&
&\omit&
&\omit&
\cr
\noalign{\hrule}
height3pt
&\omit&
&\omit&
&\omit&
\cr
&\hfil $\what{D^1}(2,1;\alpha)$ &
&\hfill $({\bf 2},{\bf {\bar 2}})$ of $SU(2)\times SU(2)$ &
&\hfil ${6k_1 k_2\over k_1 + k_2}$&
\cr
height3pt
&\omit&
&\omit&
&\omit&
\cr
\noalign{\hrule}
height3pt
&\omit&
&\omit&
&\omit&
\cr
&\hfil $\what{G}(3)$&
&\hfil ${\bf 7}$ of $G_2$&
&\hfil ${k(9k+31)\over 2(k+3)} $&
\cr
height3pt
&\omit&
&\omit&
&\omit&
\cr
\noalign{\hrule}
height3pt
&\omit&
&\omit&
&\omit&
\cr
&\hfil $\what{F}(4)$&
&\hfil ${\bf 8}$ of $Spin(7)$&
&\hfil ${2k(2k+11)\over k+4}$&
\cr
height3pt
&\omit&
&\omit&
&\omit&
\cr
}\hrule height 1.1pt
}
$$
}
\centerline{\sl Table 2: Nonlinear superconformal algebras and their central charges.}

\bigskip

For each entry we have recorded the Virasoro central charge which is
related by Jacobi identities to the level $k$ of the affine R-symmetry
algebra.   Our definition of $k$ is based on writing the current algebra OPE as
\eqn\dgz{ J^a(z) J^b(w) \sim {k \over  (z-w)^2} \delta^{ab}
+{if^{abc} \over z- w} J^c(w)~,}
in Lie algebra conventions where the Killing metric is $\delta^{ab}$ and long roots 
are normalized to have squared length two. We detail the normalizations in the 
Appendix. 

Three of the cases in the above table are actually ordinary linear superconformal algebras.  In
particular, $\widehat{OSp}(2|2;R)$ is the ${\cal N}=2$ algebra; $\widehat{SU}(1,1|2)/U(1)$ is the small
${\cal N}=4$ algebra; and $\widehat{D^1}(2,1;\alpha)$ is the large ${\cal N}=4$ algebra. 

The $\what{OSp}(4^*|2m)$ case deserves further comment.    In \FradkinKM\  
the levels of the $\what{Sp}(2m)$ and $\what{SU}(2)$ are denoted as $k_1$ and
$k_2$ respectively.    Jacobi identities relate them as $k_1 = -(k_2+2m+4)/2$.
This relation has the important consequence that $k_1$ and $k_2$ have opposite
signs for large level, clashing with unitarity. 
As explained  in the appendix,  in our conventions the $\what{SU}(2)$ level is
given by $k=k_2/2$.

For $\what{SU}(1,1|n)_{n\neq 2}$ $k$ refers to the level of the $\what{SU}(n)$ part.   The level of the
$\what{U}(1)$ (as defined in \refs{\KnizhnikWC,\BershadskyMS}) is fixed by Jacobi identities to be
$-(k+n)$.

As noted already, these nonlinear superconformal algebras arise as
the asymptotic symmetry algebras of AdS$_3$ supergravity based on
the corresponding supergroup. In particular, in \HenneauxIB\ a
classical gravity analysis was carried out yielding a nonlinear
Poisson bracket algebra.  The Poisson bracket algebras can be
understood as the large $k$ contraction of the quantum algebras,
and so the central charge that appears is given by the large $k$ limit of the
formulas appearing in Table 2.      The proportionality of
$c$ and $k$ in the classical approximation
follows directly from the fact that the classical action is proportional
to $k$.

In AdS$_3$ gravity the classical central charge is given by the
Brown-Henneaux formula $c= {{3\ell_A \over 2G_3}}$ and so $k\sim
{\ell_A \over G_3}\sim {\ell_A \over \ell_{\rm Pl}} $.  The
general central charge formulas $c(k)$ admit, for large $k$, an
expansion in $1/k$.    We now see that these corrections are
governed by powers of the Newton constant, and so represent a
series of quantum gravity corrections.  It is interesting that for
nonlinear algebras these
corrections are determined algebraically via the Jacobi
identities.  We explain in more detail how these come about from
the gravity point of view in the following two sections.

Looking at Tables 1 and 2, we can identify evidence for and against the
hypothesis that the nonlinear superconformal algebras govern the heterotic string solutions.
The $\what{OSp}(4^*|4)$ case has the best chance of being realized for
AdS$_3 \times S^2 \times T^5$,
with the $SU(2)$ R-symmetry being the isometries of the $S^2$.  It was recently
shown in \LapanJX\ that the AdS$_3\times S^2$ solution discussed in section 2
in fact has superisometry group $OSp(4^*|4)$, with the $Sp(4)$ symmetry arising
from the fermions on $T^5$. According to \jsd\ the central charge of the
supersymmetric side has the  value $c_R=-12N$.  The magnitude of this
result is desirable, although the sign is clearly not.   If we instead had
$c_R=+12N$ then from \jsd\ we would  also deduce the desired $c_L=24N$.
Note that taking $N<0$ does not fix the sign problem, as this is just a parity
transformation interchanging L and R. 

More generally, an underlying problem is that the $\what{SU}(2)$ and 
$\what{Sp}(2m)$ levels are forced by the Jacobi identities to have opposite signs, which means there are no unitary highest weight representations of this
algebra. In particular, acting with the negative level $J^a_{-1}$ on a highest weight state yields a
negative norm state. At the classical level, a negative level manifests itself by
making the energy unbounded from below. While this is obviously cause for
concern, other aspects of this proposal are sufficiently attractive to justify further
study.

This same nonunitarity problem afflicts $\what{SU}(1,1|4)$, whose $SU(4)\cong SO(6)$
is potentially the isometry group of an $S^5$.    There are again no unitary representation
since either the $\what{SU}(4)$ or $\what{U}(1)$ levels is necessarily negative.  In this case the
large $N$ central charge is $c_R=6N$, half of what we would expect, making the
identification with the heterotic worldsheet field content obscure.

There are no obvious obstacles to having unitary representations in the $\what{OSp}(8|2;R)$ and
$\what{F}(4)$ cases.  Some evidence for unitary $\what{OSp}(8|2;R)$ representations is given in \SchoutensTG.
These two cases are conceivably  related to AdS$_3 \times S^7$ and AdS$_3 \times S^6 \times S^1$.
The large $N$ central charges are $c_R = 3N$ and $c_R = 4N$ respectively,

\newsec{Quantum corrected central charges}

As we just reviewed, the nonlinear superconformal algebras imply  nontrivial relations
between the central charge and the current algebra level,
$c_R(k)$. For large $k$ we can think of this in terms of an
expansion in $1/k$.     From the bulk point of view this is an
expansion in $g_s^2$, and so the corrections correspond to quantum
corrections in the bulk.  A classical computation in the bulk is
only sensitive to the part of $c_R$ proportional to $k$.

To explain how to get the subleading terms we first consider a
simpler system consisting of a gauge field on a fixed
asymptotically AdS$_3$ geometry.  The action is\foot{In this section we work in Euclidean signature, which
accounts for the $i$ in front of the Chern-Simons term.} 
\eqn\ca{S = -{ik \over 8\pi} \int_M\! {\rm Tr}'(AdA+{2 \over 3}A^3)
+{k \over 8\pi} \int_{\p M} \!\sqrt{g}g^{\alpha\beta}{\rm Tr}'(A_\alpha
A_\beta)~, }
with $A= iA^a T^a$, $T^a$ being the generators of some group ${\cal G}$.
The notation ${\rm Tr}'$ denotes a representation independent trace,
${\rm Tr}' = {1 \over x_\rho}{\rm Tr}$, where $x_\rho$ is the Dynkin index.
As we review momentarily, this theory possesses a level $k$
${\cal G}$-current algebra. The boundary term in \ca\ should be understood in terms
of the standard holographic renormalization procedure applied to Chern-Simons
theory \KrausNB. This procedure also gives vanishing classical Brown-Henneaux
central charge, which is just a trivial consequence of the absence of
dynamical gravity (we think of taking $G_N\rightarrow \infty$ in the
Brown-Henneaux formula $c=3\ell/2G_N$).    On the other hand,
the Sugawara stress tensor for this theory has the central charge
$c={k{\rm dim}({\cal G})\over k+g}$ where $g$ is the dual Coxeter number. 
This is perfectly consistent, since
for large $k$ the Sugawara central charge is O$(1)$, and the classical
computation only sees the O$(k)$ part.  The nonzero central charge
arises from the quantum fluctuations of the gauge fields.

We first review how to derive the current algebra from the action
\ca.  The metric takes the asymptotic form
\eqn\da{ds^2 = d\eta^2 +
e^{2\eta/\ell}g^{(0)}_{\alpha\beta}dx^\alpha dx^\beta+ \ldots ~.}
We choose conformal gauge, $g^{(0)}_{\alpha\beta}dx^\alpha
dx^\beta = e^{2\omega} dw d\wb$.     In the gauge $A_\eta=0$, the
boundary conditions on the components $A_{w,\wb}$ require them to
be finite as $\eta\rightarrow \infty$.

With the boundary term in \ca, the variational principle
corresponds to holding fixed $A_\wb$ on the boundary but allowing
$A_w$ to fluctuate.  Indeed, the on-shell variation of the action is
\eqn\db{\delta S =  {k\over \pi} \int_{\p M}\! d^2w {\rm Tr}'(A_w \delta
A_\wb)~.}
The corresponding current is then
\eqn\dc{ J^a_w = -i\pi {\delta S \over \delta A^a_\wb} = ik A^a_w~.}

Under a gauge transformation
\eqn\dd{\delta A^a_\alpha = \p_\alpha \Lambda^a + f^{abc}
\Lambda^b A^c_\alpha~, }
the current varies as
\eqn\de{ \delta J^a_w = ik \p_w \Lambda^a + f^{abc}
\Lambda^b J^c_w~. }
We can now use standard CFT results (here in the context of
classical field theory on AdS) to write
\eqn\df{ \delta J^a_w(w_0) = i {\rm Res}_{w\rightarrow w_0}
\Lambda^b(w) J^b_w(w) J^a_w(w_0)~,}
from which we read off the OPE
\eqn\dg{ J^a_w(w) J^b_w(0) \sim {k \over  w^2} \delta^{ab}
+{if^{abc} \over w} J^c(0)~.  }
This confirms the existence of a level $k$ current algebra.

Now we come the question of the central charge.  This can be
computed from the stress tensor two-point function on the plane,
$\langle T_{ww}(w)T_{ww} (0)\rangle= {c \over 2 w^4}$.  Since in
\ca\ the metric only appears in the boundary term, the stress
tensor is given by%
\eqn\dh{ T_{ww} = 2\pi {\delta S \over \delta g^{ww}} = k(A^a_w)^2~.}
By the standard rules of AdS/CFT, the two-point function $\langle
T_{ww}(w)T_{ww} (0)\rangle$ is proportional to
$k^2(A^a_w(w))^2(A^a_w(0))^2$, where at the classical level $A^a_w$
is  the gauge field consistent with the equations of motion and
boundary conditions.   In computing the stress tensor correlator
on the plane we have the boundary condition $A^a_{\wb}=0$ and
hence the classical solution has $A^a_w=0$ as well.  We conclude
that the stress tensor correlator vanishes classically, and then so too does
the classical central charge.

What makes the quantum central charge nonvanishing is that the gauge field
$A^a_w$ undergoes quantum fluctuations controlled by $1/k$.   The
relevant fluctuations are localized at the AdS boundary, since
this is where the stress tensor lives.   To quantize these
fluctuations we can use the well known fact that bulk Chern-Simons
theory localizes to a WZW theory on the boundary
\refs{\WittenHF,\MooreYH,\ElitzurNR}.

To proceed (see {\it e.g.} \CarlipZN) we write
\eqn\ec{ A = g^{-1}dg + g^{-1} \Ab g~. }
Here $\Ab$ is a background connection with the prescribed boundary
condition for $A_\wb$. Substituting into \ca\ we get, in conformal
gauge,
\eqn\ed{\eqalign{S_{gauge} &= -{ik \over 8\pi} \int_M\! {\rm
Tr}'(AdA+{2 \over 3}A^3) +{k \over 8\pi} \int_{\p M}
\!\sqrt{g}g^{\alpha\beta}{\rm Tr}'(A_\alpha A_\beta) \cr & =-{ik \over
8\pi} \int_{\p M} \! {\rm Tr}'( g^{-1}\p g g^{-1} \pb g^{-1}-2
g^{-1} \p g \Ab_{\wb} )-{ik \over 24\pi} \int_M {\rm Tr}'( g^{-1}
dg)^3~.}}
That is to say, we get a WZW model with the current $J\sim
kg^{-1}\p g$ coupled to the external potential $\Ab_\wb$.

In this theory the conformal boundary metric
$g^{(0)}_{\alpha\beta}$ couples to the Sugawara stress energy
tensor, which is the quantum version of the classical formula \dh.
Computation of the AdS$_3$ central charge from the stress tensor
two-point function now becomes the same computation as in the
boundary WZW model.   Hence, we automotically recover the
Sugawara central charge $c={k{\rm dim}({\cal G})\over k+g}$. The
point we wish to make here is that the full result follows from a systematic
application of the AdS/CFT correspondence.

\newsec{Nonlinear superconformal algebra from holographic renormalization}

The nonlinear superconformal algebras have a ``physical" realization as
the asymptotic symmetry algebras of AdS$_3$ supergravities \refs{\deBoerIP,\ItoVD,\HenneauxIB}.    In  \HenneauxIB\ the algebras were found by applying the
Regge-Teitelboim method  \ReggeZD.  In this section we will reproduce the
result in the framework of holographic renormalization, which is the most
convenient  approach within AdS/CFT.  The logic is exactly the same as 
led to the current algebra \dg.    In this approach, the Sugawara
stress tensor contribution that was added by hand in \HenneauxIB\ arises
automatically, see \dh.

The novelty of the nonlinear superconformal algebras lies in the OPE of
two supercurrents, which has the schematic structure
\eqn\zza{ G(w)G(0) \sim  {1\over w^3} + {J \over w^2} +{\partial J \over w}
+ {T \over  w} +{  JJ \over w}~.}
 In particular, the nonlinearity referes to the appearance of $JJ$.   In
 the following we will derive \zza\  in the simplified case where  the metric is pure AdS$_3$.   
 Including a  general metric is completely straightforward, but clutters the computation.

The  bulk supergravity
action contains the usual Einstein-Hilbert term with negative
cosmological constant, Chern-Simons terms for the gauge-fields, and
the most relevant piece for the present purposes, the Rarita-Schwinger term for the gravitino
\eqn\ca{S_{RS} = {i \over 16\pi G_N}\int\! d^3x \left(
\epsilon^{MNP}{\psib}^i_M \Dc^{ij}_N \psi^j_P + {~e\over 2\ell}
\psib^i_M \Gamma^{MN}\psi^i_N\right)~. }
Here $i=1\ldots~{dim}(\cal{R})$, where $\cal{R}$ is the
representation of the gauge group under which the gravitinos
transform.  The covariant derivative is defined as
\eqn\cb{ \Dc^{ij}_M= \left(\p_M +{1\over 4}
\omega_M^{\Ah\Bh}\Gamma_{\Ah\Bh}\right)\delta^{ij} + A^a_M
(T^a)^{ij}~,}
where $a=1,\ldots,dim({\cal G})$ and $T^a$ are generators of the
gauge group $\cal G$.  Three-dimensional indices are capitalized
$M,N,P$, with tangent space indices being indicated with hats. Later,
we will need indices on the two dimensional boundary labeled by
lower-case letters $m,n$. 

The OPE \zza\ will be extracted from the supersymmetry transformations of the gravitinos,
which are
\eqn\zi{\delta_\epsilon \psi^i_M=\left(\Dc_M\epsilon+{1\over
2\ell}e^\Ah_M\Gamma_\Ah\epsilon \right)^i~.}

\subsec{Deriving the  boundary supercurrent}

To derive the boundary supercurrent we 
must first obtain the leading radial behaviour of the gravitino from its
equation of motion.   As noted above, we take the metric to be pure AdS$_3$
\eqn\cc{ds^2 = d\eta^2 + e^{2\eta/\ell}dw d\wb~.}
We work in the gauge
$\psi^i_\eta=A_\eta=0$, for which  the equations of motion
\eqn\cf{\epsilon^{MNP} \Dc^{ij}_N \psi^j_P + {e\over
2\ell}\Gamma^{MN}\psi^i_N =0~,}
simplify to
\eqn\ze{\eqalign{0&=\left(\partial_\etah+{1\over
2\ell}\Gamma_\etah\right)\psi^i_w~, \cr
0&=\left(\partial_\etah+{1\over
2\ell}\Gamma_\etah\right)\psi^i_\wb~.}}
The main simplification here is that  the covariant derivative in the radial $\eta$ direction
has vanishing spin-connection\foot{The spin-connection
$\omega^{ab}_\eta$ only receives a contribution from gravitino
induced torsion which will vanish at the boundary  $\eta
\rightarrow\infty$.} and so becomes $\Dc_\eta=\partial_\eta=\partial_\etah$.
Our convention for the antisymmetric tensor is $\epsilon_{\wh\wbh\etah}=+1$ and
we choose $\Gamma_{\Ah\Bh\Ch}=-\epsilon_{\Ah\Bh\Ch}$. Since
$\Gamma^{\wh\wbh}=-\Gamma^{\wbh\wh}=+\Gamma_\etah$ we have
$\Gamma_\etah^2=1$, {\it i.e.}, $\Gamma_\etah$ is the 2d
chirality operator.  Decomposing the spinors as
\eqn\zf{\Gamma_\etah\psi^i_{m\pm}=\pm\psi^i_{m\pm}~,}
we can solve \ze\ with
\eqn\zg{\eqalign{\psi^i_{w}&=\psi^i_{-w}e^{\eta/2\ell}+\psi^i_{+w}e^{-\eta/2\ell}~,
\cr\psi^i_{\wb}&=\psi^i_{-\wb}e^{\eta/2\ell}+\psi^i_{+\wb}e^{-\eta/2\ell}~.}}
The leading boundary components $\psi^i_{-,w,\wb}$ of both
spinors have negative 2d  chirality.

As usual in AdS/CFT, the leading boundary component plays the role of a
source in the CFT coupled to a boundary current (in this case the supercurrent). 
To extract the current we now consider the
on-shell variation of the action \ca\
\eqn\ck{\eqalign{\delta S_{RS} &= {i\over 16\pi
G_N}\int\!d^2w~\epsilon^{mn} \left(\psib^i_m \delta
\psi^i_n\right)\cr&={i\over 16\pi G_N}\int\!d^2w ~\left(\psib^i_w
\delta \psi^i_\wb -\psib^i_\wb \delta \psi^i_w \right) \cr &=
{i\over 16\pi G_N}\int\!d^2w~\left(\psi^i_{+w} \delta
\psi^i_{-\wb}-\psi^i_{-w} \delta \psi^i_{+\wb} -\psi^i_{+\wb} \delta
\psi^i_{-w} +\psi^i_{-\wb} \delta \psi^i_{+w} \right), }}
where the Majorana conjugate spinor is defined by $\psib=\psi^T C$
with  charge conjugation matrix $C= i\sigma^2$. 

We seek a  variational principle  in which we hold fixed the leading
boundary components.  For this to be valid we need to add a boundary
term to the action to cancel the unwanted variations in \ck, 
\eqn\cl{\eqalign{ S_{bndy}&= {i\over 16\pi G_N} \int\!d^2x ~
\epsilon^{mn} ~\psi^i_{+m}\psi^i_{-n} \cr &= {i\over 16\pi G_N} \int\!
d^2w \left( \psi^i_{+w}\psi^i_{-\wb}-\psi^i_{+\wb}\psi^i_{-w}\right)~,
}}
leaving the desired result
\eqn\cm{ \delta (S_{RS}+S_{bndy}) ={i\over 8\pi
G_N}\int\!d^2w~\left(\psi^i_{+w} \delta \psi^i_{-\wb} -\psi^i_{+\wb}
\delta \psi^i_{-w} \right)~.}
We now define the boundary supercurrent via
\eqn\cn{\delta S \equiv  {1 \over 8\pi} \int\!d^2w~
\overline{G}^{im}\delta\psi^i_m~,}
where the prefactor was fixed for later convenience. 
We thus  find the holomorphic and anti-holomorphic boundary  supercurrents
\eqn\co{ G^i_{+w}\equiv G^i={i\over 2 G_N} \psi^i_{+w}~,\quad
G^i_{+\wb}\equiv \overline{G}^i=-{i\over 2 G_N} \psi^i_{+\wb}~.
}
Although we obtain both holomorphic and antiholomorphic
supercurrents, we will see that the action \ca\
only provides the boundary OPE of the holomorphic side.  The 
corresponding antiholomorphic boundary algebra is obtained 
from the additional Rarita-Schwinger term 
for a gravitino $\psi^{\tilde{i}}$ which transforms in a
representation $\widetilde{\cal{R}}$ and with opposite sign AdS$_3$
mass term $-{1\over
2\ell}\psib^{\tilde{i}}_M\Gamma^{MN}\psi^{\tilde{i}}$.

\subsec{Bulk symmetries and the boundary OPE}

The Noether theorem relates the OPE between the supercurrent $G^i$ and
another current $\Phi$  to the variation of the supercurrent
$\delta G^i$ under the symmetry transformation generated by $\Phi$.
Here we are interested in the OPE between two supercurrents, which is
then related to the supersymmetry transformation by the standard CFT expression
\eqn\zr{ \delta_{\epsilon} G^i_w(w_0) = i {\rm Res}_{w\rightarrow
w_0} \epsilon^j(w) G^j_w(w) G^i_w(w_0)~,}
where the expression for $\delta_{\epsilon} G^i_w$
is that induced by the susy transformation \zi.

According to the AdS/CFT dictionary, the two-point function on the right hand side
is
 \eqn\zm{G^iG^j=\left.{\delta^2 S\over\delta\psi^i_{-\wb}\delta\psi^j_{-\wb}}\right\vert_{
 \psi^j_{-w}~=~\psi^j_{-\wb}~=~0}}
The boundary conditions $ \psi^i_{-w}= \psi^i_{-\wb}=0$ ensure that the
external sources  are turned off.

Our susy transformations should preserve our gauge choice $\psi^i_\eta=0$, so we need
\eqn\zj{\delta_\epsilon \psi^i_\eta\equiv
0=\left(\partial_\etah+{1\over 2\ell}\Gamma_\etah\right)\epsilon^i~,}
which determines the radial dependence of the spinorial parameter as
\eqn\zk{\epsilon^i=\epsilon^i_{-}e^{\eta/2\ell}+\epsilon^i_{+}e^{-\eta/2\ell},}
by repeating the manipulations yielding  \ze--\zg.

A symmetry transformation should also be one that leaves the sources invariant, so
we need $\delta \psi^i_{-,w,\wb}=0$.  This condition  relates
$\epsilon^i_+$ and $\epsilon^i_{-}$ as follows.  We need the covariant derivatives
\eqn\zm{\Dc_w+{1\over 2\ell}e^\Ah_w\Gamma_\Ah=\p_w+A_w
+{1\over\ell}e^\wh_w\Gamma_\wh~,\quad \Dc_\wb+{1\over
2\ell}e^\Ah_\wb\Gamma_\Ah=\p_\wb+A_\wb~,}
and a representation of the Gamma matrices satisfying our
conventions
\eqn\cib{\Gamma_\etah = \left(\matrix{1&0\cr0&-1}\right),~\quad
\Gamma_{\wh} = \left(\matrix{0&0\cr -\sqrt{2}&0}\right)~,\quad
\Gamma_{\wbh} = \left(\matrix{0&-\sqrt{2}\cr0&0}\right)~.}
Then the remaining bulk supersymmetry transformations read
\eqn\ct{\eqalign{\delta \psi^i_w  &= \left[
(\p_w+A_w)\left(\matrix{1&0\cr0&1}\right) +
{1\over\ell\sqrt{2}}\left(\matrix{0&0\cr -\sqrt{2} & 0}\right)
\right]^i_j\epsilon^j ~,\cr
\delta \psi^i_\wb  &= \left[
(\p_\wb+A_\wb)\left(\matrix{1&0\cr0&1}\right)
\right]^i_j\epsilon^j~.}}
The constraint  $\delta \psi^i_{-w}=0$
now determines $\epsilon_+$ as
\eqn\zn{\epsilon^i_+=\ell\left(\p_w+A_w\right)^i_j\epsilon^j_{-}~,}
and $\delta\psi^i_{-\wb}=0$ requires
\eqn\zo{\left(\p_\wb+A_\wb\right)^i_j\epsilon^j_{-}=0~.}
From \zo\ we find that the supersymmetry parameter is covariantly
holomorphic, as expected for the holomorphic part of the
boundary algebra. Collecting results, we find the variation of the
supercurrent
\eqn\zq{\eqalign{\delta G^i&= {i\over 2 G_N}\delta
\psi^i_{+w}\cr &= {i\over 2
G_N}\left(\p_w+A_w\right)^i_j\epsilon^j_{+}\cr &={i\ell\over 2
G_N}\left(\p_w+A_w\right)^i_j\left(\p_w+A_w\right)^j_k\epsilon^k_{-}~.
}}
We can now read off the boundary OPE of two supercurrents by
using \zr\ and the
expression \dc\ for the boundary  current $J^a_w=ik A^a_w$.
We find
\eqn\zs{G^i(w)G^j(0)\sim{\ell\over G_N}\left({\delta^{ij}\over
w^3}+{1\over k}{J^a (T^a)^{ij}\over w^2}+{1\over 2k}{\p_w J^a_w
(T^a)^{ij}\over w}+{1\over 2 k^2}{J^a_w
J^b_w(T^a)^{ik}(T^b)_k^{~j}\over w}\right)~.}
To compare with the CFT literature we need to trade $\ell$ for $k$.  
We first use the Brown-Henneaux formula $c= 3\ell/2G_N$.  For the 
simple supergroups with irreducible $\rho$  (the cases $OSp(m|2;R), G(3)$ and
$F(4)$) the large $k$ relation between $c$ and $k$ is\foot{There are analogous
case-by-case formulas for the other groups.}  
\eqn\zsa{ c ={3k \over 2\chi}~,\quad \chi ={x_\rho { \rm dim}({\cal G})\over {\rm dim}(\rho)({\rm dim}(\rho)-1)}~,}  
as can be verified from Table 2. This gives     
\eqn\zt{G^i(w)G^j(0)\sim  {k\over \chi} {\delta^{ij}\over w^3}+{1 \over \chi} { J^a
(T^a)^{ij}\over w^2}+{1 \over 2\chi}{\p_w J^a_w (T^a)^{ij}\over
w}+{1 \over 2 \chi k} { J^a_w
J^b_w(T^a)^{ik}(T^b)_k^{~j}\over w}~.}
The term quadratic in the currents includes the Sugawara contribution to the stress-tensor 
(see \dh) so we can write the complete expansion as
\eqn\zta{G^i(w)G^j(0)\sim  {k\over \chi} {\delta^{ij}\over w^3}+{1 \over \chi} { J^a
(T^a)^{ij}\over w^2}+{2T\delta^{ij}\over w} + {1 \over 2\chi}{\p_w J^a_w T^a\over
w}+{1 \over 2 \chi k} { J^a_w
J^b_w P_{ab}^{ij}\over w}~,}
where $P_{ab}^{ij} = \half\{ T^a , T^b \}^{ij} -2 \chi \delta_{ab}\delta^{ij}$. This is our final result
for the OPE of two supercurrents. For specific groups the Lie algebra may be such that 
$P_{ab}^{ij}=0$ identically, and so the nonlinear term in \zta\ vanishes. This happens
for $SO(3)$, corresponding to the usual ${\cal N}=4$ SCA. But for more general
groups  the nonlinearity persists. The coefficients of all the  terms in \zta\ agree
for large $k$ with those previously determined by analyzing the Jacobi identities of the
quantum nonlinear SCA (see {\it e.g.},  \BinaBM.) 

As we have emphasized, the computation done here is classical and so our expressions
are valid only up to corrections suppressed by ${\cal O}(1/k)$.  For instance, in the
quantum treatment the $JJ$ product in \zt\ requires normal ordering.   On the one
hand, the $1/k$ corrections can be determined algebraically by demanding a consistent
operator algebra obeying the Jacobi identities.  On the other hand, we can also
understand the origin of the $1/k$ corrections from the gravity point of view, in 
the same spirit as discussed at the end of section 5.   The corrections come from
quantizing the fluctuations around AdS, and in particular the pure gauge modes
localized near the boundary.  In section 5 these modes were described by a 
WZW model, while in the present case they are described by a super Liouville
theory, as shown in \HenneauxIB.    Quantizing this theory will then yield the
full central charge expressions of Table 2.

\newsec{String Theory on AdS$_3\times S^2$}
In this section we make some comments on the heterotic $\sigma$-model
with AdS$_3\times S^2$ target space. This is the holographic dual of $N$
fundamental heterotic strings.

Let us begin by recalling the basics of heterotic string theory on a
$SL(2)_k\times SU(2)_{k^\prime}\times U(1)^4$ target space. This is just a sum
of $WZW$-models. Keeping track of just the bosons, the world-sheet central charge
becomes
\eqn\fa{
c_{B,{\rm ws}} = {3k_B\over k_B + 2} +  {3k^\prime_B\over k^\prime_B - 2}
+4~.
}
To get the critical central charge we need $c_{B, {\rm ws}}=10$,
as for 10 free bosons. This gives the condition
\eqn\fb{
k_B=k^\prime_B+4~.
}
Of course the heterotic model also has additional left-moving field such that
the total left-moving central charge has the correct value $c_{L,{\rm ws}}=26$.

The right-moving fermions in the heterotic model change the accounting in two ways.
First, their central charge contribute $c_{F,{\rm ws}}=5$ such that the total right-moving
central charge has the correct value $c_{R,{\rm ws}}=15$. Second, world-sheet
supersymmetry demands that these fermions are organized into
$\widehat{SL}(2)\times \widehat{SU}(2)\times \widehat{U}(1)^4$ current algebra. They contribute
$k_F =-2$, $k_F^\prime =2$ such that the total levels on the right side become
$k_{R,{\rm tot}} = k_B -2$, $k^\prime_{R,{\rm tot}} = k^\prime_B -2$.

In order to fully specify the model we must find $k_B$, which then determined
the level of all the world-sheet current algebras.
String theory on AdS$_3$ \refs{\GiveonNS,\KutasovXU} has a spacetime Virasoro
algebra with central charge related to the level of the world-sheet $\widehat{SL(2)}$ current
as
\eqn\fc{
c_{\rm spacetime} = 6k_{\rm tot}N~.
}
The overall factor of $N$ is due to winding of the map between string world-sheet and
spacetime target. 
The general construction above describes a bound state of fundamental strings together
with NS fivebranes. We now try to get rid of the fivebranes.  Given only fundamental strings, our expectation from spacetime considerations is $c_R=12N$,
$c_L=24N$ (with $1/N$ corrections coming from string loop corrections). Then \fc\ indicates $k_B=k_{L,{\rm tot}}=4$ by considering the left-movers.
This fixes the right-moving level to $k_{R,{\rm tot}} = k_B -2=2$ and then \fc\ gives
the correct right-moving central charge as well. This result was not automatic so it gives
a modest check on the basic accounting.

As we have discussed, the nonlinear superconformal algebras determine the
corrections of \fc\ due to string loops. In principle such corrections
could be verified by direct computation of string loops in the $\sigma$-model.
Alhough it is unlikely that such explicit computations are ever going to be practical
it is meaningful that there is a concrete and nontrivial prediction.

We next consider the $SU(2)$ factor of the world-sheet theory. According
to \fa\ with $k_B=4$ the bosonic $SU(2)$ level becomes $k^\prime_B = 0$. All that remains
is then $k^\prime_{R}=2$ from the right-moving fermions. The symmetry is
therefore reduced, from $SU(2)_R\times SU(2)_L$ in a generic $\sigma$-model
to $SU(2)_R$. This is consistent with describing a spacetime $S^2$ rather than
$S^3$. In the present context this is just what we want. The disappearance of one
of the bosonic $SU(2)$'s is reminiscent of the model for fundamental strings presented
in \GiveonPR.

According to \refs{\GiveonNS,\KutasovXU} a world-sheet current algebra at
level $k_{\rm ws}$ gives rise in spacetime to an affine current with the level
\eqn\fd{
k_{\rm spacetime} = k_{\rm ws}N~.
}
In the present case, the world-sheet $\widehat{SU}(2)$ with level $k^\prime_{R}=2$
gives a spacetime $\widehat{SU}(2)$ with level $2N$, in agreement with \jsd.
It is also worth noting that $\widehat{SU}(2)_2$ can be bosonized to a supersymmetric
to a supersymmetric $\widehat{U}(1)$ with the boson at the self-dual radius. This is
the value \aak\ of the fifth circle that appears in the classical geometry.

As an aside, we make the following suggestive observation. Since
 all the right-moving world-sheet fermions are free they
form an $\widehat{SO}(8)$ at level 1, as is familiar from strings in flat space. According
to \fd\ this would give a spacetime $\widehat{SO}(8)$ at level $k_{SO}(8) = N$. Since we
would like to 
compactify five of the bosonic directions (although our construction only has a manifest
$T^4$)  the spectrum generically respects only the
$\widehat{SO}(3)_1\subset \widehat{SO}(8)_1$
subgroup, which appeared above as an $\widehat{SU}(2)$.
At some points in moduli space the spectrum respects the centralizer as well, an
$\widehat{SO}(5)$ at level 1. According to \fd\ this gives a spacetime
$\widehat{SO}(5)$ at level $N$.

The direct construction of the superisometry  was given in \LapanJX. The
corresponding nonlinear algebra is $\what{OSp}(4^*|4)$. This algebra has spacetime
R-symmetry $\what{SO}(3)_{-2N}\times\widehat{SO}(5)_N$. It is not clear what one
is to make of the negative level, and the related fact that the representations appearing in the
spacetime spectrum are nonunitary. It is therefore also unclear whether one should
take seriously the apparent match between (the absolute value of) these levels and
those discussed in the previous paragraph.

So far we focussed on the bosonic symmetries and found promising results using simple
and rather robust arguments. We next consider supersymmetry which will turn out to be
more confusing. The world-sheet theory must of course respect the right-moving
supersymmetry because it is gauged. Having introduced a bosonic
$\widehat{SL}(2)_4$ the supercurrent must be appropriate for $\widehat{SL}(2)$
as well. In the present context we can take
\eqn\fe{
T_F =  \eta_{AB} \psi^A j^B -  {i\over 6}\epsilon_{ABC}\psi^A \psi^B \psi^C
- {i\over 6}\epsilon_{A'B'C'}\chi^{A'} \chi^{B'} \chi^{C'} + \lambda^i \partial Y_i~,
}
where $j^B$ denote the bosonic $\widehat{SL}(2)$ currents, $\psi^A$ are the $SL(2)$ fermions,
$\chi^{A'}$ are three fermions forming $\widehat{SU}(2)_2$, and $(\lambda^i,Y_i)$ realize the
supersymmetric $\widehat{U}(1)^4$.
Bosonizing as usual the $10$ fermions into $5$ $H_I$'s, there are $32$ candidate
spacetime supersymmetries
\eqn\ff{
Q_\alpha = \exp\left(\half i \sum_{I=1}^5\epsilon_I H^I\right)~.
}
Mutual locality imposes the GSO projection $\prod_{I=1}^5\epsilon^I=1$ and locality
with respect to the world-sheet supercurrent \fe\ further imposes
$\prod_{I=1}^3\epsilon^I = 1$. Therefore, there are only $8$ spacetime supersymmetries
whereas we expected enhancement to $16$ supersymmetries. In fact, it is this
enhancement that forces the appearance of a the nonlinear superconformal algebra in
spacetime, our main interest. The key test for a successful $\sigma$-model is to
achieve the correct spacetime supersymmetry.

There is one more important ingredient to consider: the construction may need
specific discrete identifications realized by some orbifold. Generally $\sigma$-models
realize $S^2$ as the Lenz space $S^3/Z_p$ by taking an asymmetric $Z_p$ orbifold
of $SU(2)$. The world-sheet central charges discussed above are not affected by such an
orbifold, but other symmetries including supersymmetries depend sensitively on such
discrete choices. An explicit model in the context of string theory in
AdS$_3$ \refs{\GiveonNS,\KutasovXU} was studied in \KutasovZH. In this model the
currents are invariant under the orbifold at the lowest level $p=2$ and so the model
is precisely the one discussed above. This model appears to have only $8$
spacetime supersymmetries and so it is not quite the correct dual.

Another approach to the asymmetric orbifold $S^3/Z_p$ \GiddingsWN\ represents
some of the left-movers as fermions and use these to balance the anomaly from the
asymmetric gauging of the $SU(2)$ (some useful details are given in \JohnsonJW).
In this model there is an enhanced discrete symmetry which is not manifest when
the left-movers are represented as bosons. The special case where the left-moving
fermions are neutral under the gauging (so their charges $Q=0$) was proposed in \LapanJX\ as the holographic dual of fundamental strings.
However, it is (again) not clear how to achieve the correct spacetime
supersymmetry.

\newsec{Discussion}

In this paper we studied aspects of the holographic description of fundamental heterotic
strings, and in particular  the hypothesis that they are governed by nonlinear superconformal algebras.
There are many open questions and further avenues to pursue; we close by mentioning a few.

There are several arguments supporting the  appearance of $\widehat{OSp}(4^*|4)$ in the five dimensional heterotic string.   In particular, the identification in \LapanJX\ of $OSp(4^*|4)$ as the superisometry
group is an excellent clue.  It should be possible to actually prove this assertion, at least in the
context of five dimensional $R^2$ supergravity, by constructing explicitly the generators of the
algebra.   Note that this is different than finding the superisometry group; indeed, the structure of
the full nonlinear algebra implies that the Lie algebra of $OSp(4^*|4)$ is not in fact a subalgebra of the full $\widehat{OSp}(4^*|4)$, except in the $k\rightarrow \infty$ limit.

Assuming that the nonlinear algebras indeed appear in the present context, to fill in the CFT
side of the AdS/CFT correspondence we need to identify  boundary CFTs possessing these symmetry algebras.
Not much is known about such field theories.

In \WittenKT\  Witten gave a proposal for the boundary CFT description of pure gravity in
AdS$_3$.  It might similarly be worthwhile to consider the CFT dual of pure AdS$_3$ supergravity
based on the various supergroups.   As we have emphasized here, the Jacobi identities by themselves
already lead to highly nontrivial quantum gravity  predictions for corrections to the black hole entropy.
Pursuing the logic of \WittenKT\ should lead to further structure.

\bigskip
\noindent {\bf Acknowledgements:} \medskip \noindent

\medskip
We thank A. Dabholkar, J. Davis, E. D'Hoker, D. Marolf, A. Sen, A. Strominger and E. Witten
for discussions. FL thanks CERN for hospitality as the write-up was completed.
The work of PK is supported in part by NSF grant PHY-0456200.
The work of FL is supported  by DOE under grant DE-FG02-95ER40899. The 
work of AS is supported by an NSF IGERT Fellowship. 

\appendix{A}{Conventions and normalizations}

The detailed expressions for the nonlinear superconformal algebras depend on many 
group theory conventions, which can differ among the cited references \refs{\KnizhnikWC-\BowcockBM}.  
For convenience, here we give our conventions and explain the relation of
our formulas to those in the references.  Our primary reference is \yellow.

\subsec{Basic Lie algebra} 

The scalar product of two Lie algebra elements $X$ and $Y$ is given by the 
basis independent Killing form
\eqn\ka{ K(X,Y)= {1 \over 2g} \Tr(ad X ad Y)~,}
where $g$ is the dual Coxeter number, defined below. We normalize the 
generators of the Lie algebra such that $K(T^a,T^b) = \delta^{ab}$. 
Roots are introduced by writing the algebra in its standard Cartan form
\eqn\kb{ [H^i, E^\alpha] = \alpha^i E^\alpha~,}
with $H^i$ properly normalized. We take the long roots of the algebra 
to have length squared $\psi^2=2$. 

All other normalizations are fixed by these conventions. The generators
in some general representation $\rho$ are normalized according to
\eqn\kd{ \Tr_\rho (T^a T^b) = 2 x_\rho \delta^{ab}~,}
which defines the Dynkin index $x_\rho$.  
The dual Coxeter number is 
\eqn\kda{
g=x_{\rm adj} = \half C_2({\rm adj})~,
}
where the quadratic Casimir in the adjoint representation is defined by 
\eqn\kdb{
f^{acd}f^{bcd}=C_2({\rm adj})\delta^{ab}~.
}
In these  conventions we have:
\bigskip
\vbox{
$$\vbox{\offinterlineskip
\hrule height 1.1pt
\halign{&\vrule width 1.1pt#
&\strut\quad#\hfil\quad&
\vrule width 1.1pt#
&\strut\quad#\hfil\quad&
\vrule width 1.1pt#
&\strut\quad#\hfil\quad&
\vrule width 1.1pt#\cr
height3pt
&\omit&
&\omit&
&\omit&
\cr
&\hfil Algebra&
&\hfil Dual Coxeter $g$ &
&\hfil Dynkin index of defining rep. &
\cr
height3pt
&\omit&
&\omit&
&\omit&
\cr
\noalign{\hrule height 1.1pt}
height3pt
&\omit&
&\omit&
&\omit&
\cr
&\hfil $SU(n)$ &
& \hfil n&
&\hfil ${1\over 2}$&
\cr
height3pt
&\omit&
&\omit&
&\omit&
\cr
\noalign{\hrule}
height3pt
&\omit&
&\omit&
&\omit&
\cr
&\hfil $SO(n>3)$ &
& \hfil n-2&
&\hfil $1$&
\cr
height3pt
&\omit&
&\omit&
&\omit&
\cr
\noalign{\hrule}
height3pt
&\omit&
&\omit&
&\omit&
\cr
&\hfil $Sp(2n)$ &
& \hfil n+1&
&\hfil ${1\over 2}$&
\cr
}\hrule height 1.1pt
}
$$
}
\centerline{\sl Table 3: basic normalizations of some important Lie algebras}

In the important case of $SO(3)\simeq SU(2)$ we cannot use $n=3$ in $SO(n)$ 
because precisely for this case the long root disappears, and so the normalization 
is off.  To compute the Dynkin index of the vector representation of $SO(3)$
we consider instead the adjoint of $SU(2)$ and so get 
$x_{v}( SO(3)) = x_{\rm adj}( SU(2)) = g(SU(2)) = 2$.  

\subsec{Conventions for WZW models}
We follow \yellow\  and write the topological term in the WZW model as 
\eqn\kaa{  k\Gamma = {-ik \over 24\pi x_\rho} \int\!\Tr_\rho [(g^{-1} dg)^3 ]~, }
where $\Tr_\rho$ represents the trace in the representation $\rho$ and the
Dynkin index was introduced in \kd. The factor of $x_\rho$ in the 
denominator of \kaa\ makes $k\Gamma$ independent of the choice
of $\rho$.  According to \yellow\ the WZW model with our normalizations
has the current algebra OPE
\eqn\kac{J^a(z) J^b(0) \sim {k \delta^{ab} \over z^2} + i f^{abc} {J^c(0) \over z}~, 
}
which is equivalent to the algebra
\eqn\kad{
[J^a_n,J^b_m]= if^{abc} J^c_{n+m} +kn \delta^{ab} \delta_{n+m}~. 
}
For the action to be well defined the level $k$ must be an integer for any group; for $SO(3)$ it must be an 
even integer. When comparing \kac , \kad\ with the literature it is essential
to normalize the currents consistent with \kda, \kdb, {\it i.e.} such
that
$f^{ade}f^{bde} =C_2({\rm adj})\delta^{ab}=2g\delta^{ab}$. 

\subsec{Central charges of nonlinear algebras}

In order to read off the correct results for the central charges of the nonlinear 
SCAs we must apply the normalizations above carefully. Some examples:

\vskip .2cm
\noindent
{\bf $\what{OSp}(n|2:R)$}
\vskip .2cm

Knizhnik \KnizhnikWC\ uses $f^{abc}f^{abd} = 2(n-2)\delta^{cd}=2g_{SO(n)} \delta^{cd}$
and writes the OPE as \kac, albeit with $k\to S$. This agrees with our normalizations
so $S_{\rm there}=k_{\rm here}$ and then  \KnizhnikWC\ gives
\eqn\ma{c={k(6k+n^2-10)\over 2(k+n-3)}= 3k+\ldots~. }

\vskip .2cm
\noindent
{\bf $\what{SU}(1,1|n)_{n\neq 2}$}
\vskip .2cm

In this case Knizhnik \KnizhnikWC\ writes $f^{abc}f^{abd} = 4N\delta^{cd}=4g_{SU(N)} \delta^{cd}$ and writes the OPE as \kac, again with $k\to S$. Here the normalizations
are off so that $S_{\rm there}=2k_{\rm here}$. In our notation \KnizhnikWC\ gives
\eqn\mb{c={3k(2k+n)+(n-1)(1+(n+1)k) \over k+n-1 }= 6k+\ldots~. }

\vskip .2cm
\noindent
{\bf $\what{F}(4)$}
\vskip .2cm

We use  \FradkinKM.   Taking $\psi^2=2$ all normalizations agree, so 
$k_{\rm there}=k_{\rm here}$. Then 
\eqn\mc{c={2k(2k+11)\over k+4} = 4k+\ldots~. }

\vskip .2cm
\noindent
{\bf $\what{G}(3)$}
\vskip .2cm

We use  \FradkinKM.   Taking $\psi^2=2$ all normalizations agree again, so 
$k_{\rm there}=k_{\rm here}$. Then
\eqn\md{c={k(9k+31)\over 2(k+3)} = {9 k\over 2}+\ldots~. }

\vskip .2cm
\noindent
{\bf $\what{SU}(1,1|2)/U(1)$}
\vskip .2cm

This is the standard ${\cal N}=4$ SCA for which 
\eqn\me{c=6k}

\vskip .2cm
\noindent
{\bf $\what{D^1}(2,1;\alpha)$}
\vskip .2cm

This is the ``large" ${\cal N}=4$ algebra. Our conventions agree with {\it e.g.} \GukovFH\
so that $k_{\rm there}=k_{\rm here}$. The result is 
\eqn\mf{c={6k_1 k_2 \over k_1+k_2} }

\vskip .2cm
\noindent
{\bf $\what{OSp}(4^*|2m)$}
\vskip .2cm

We  use \FradkinKM.  We label $\what{SU}(2)$ currents as
$J^{\alpha\beta}$ and $\what{Sp}(2m)$ currents as $J^{AB}$. Both are
symmetric (we're using Sp convention for $SU(2)$).  The 
affine $\what{SU}(2)$
algebra is
\eqn\wg{[J_m^{\alpha\beta},J_n^{\gamma\delta}]=
\epsilon^{\beta\gamma}
J^{\alpha\delta}_{m+n}+\epsilon^{\alpha\delta}
J^{\beta\gamma}_{m+n}+\epsilon^{\alpha\gamma}
J^{\beta\delta}_{m+n}+\epsilon^{\beta\delta} J^{\alpha\gamma}_{m+n}
-k_2(\epsilon^{\alpha\gamma}\epsilon^{\beta\delta}+\epsilon^{\beta\gamma}\epsilon^{\alpha\delta})m
\delta_{m+n}~, }
and the analogous structure for the $\what{Sp}(2m)$ currents with their
level being $k_1$.  Jacobi identities determine the central charge as
\eqn\wh{c= -{3k_2(k_2+2m+4) \over k_2-2m+4}
+{6k_2+(2m+1)(m-2)(k_2+2m+4)\over k_2-2m+4}~,}
and further relate the levels as
\eqn\wi{ 2k_1+k_2+2m+4 =0~.}
For large level the relative signs of $k_{1,2}$ must be opposite. 
To get a positive central charge we need $k_2<0$.  

We need to write the $\what{SU}(2)$ level $k_2$ in our conventions. 
The commutation relations \wg\ give
\eqn\wj{ [J^{12}_0, J^{11}_0  \pm J^{22}_0 ] =- 2 (J^{11}_0 \mp J^{22}_0)  ~, }
which identifies $J^3 = {1 \over \sqrt{2} } J^{12}$ as the properly normalized 
$SU(2)$ Cartan generator which gives roots of length squared $2$.   
Then \wh\ gives
\eqn\wl{ [J^3_n, J^3_m ] = {k_2 \over 2} m \delta_{m+n}~,}
and comparison with \kad\ identifies the level in our conventions
as $k_{\rm here} = \half k_2$. 

\listrefs 

\end